Review Article

# Cultural and Regional Influences on Global AI Apprehension

**Victor Frimpong**[1]

1. Swiss Business School Zurich (SBS), Switzerland

**The discussion surrounding artificial intelligence (AI) is primarily influenced by Western viewpoints, often neglecting the cultural, regional, and socio-political factors that affect global perceptions of AI. This study examines how these factors shape AI apprehension in various regions, including Africa, Asia, Latin America, and the Middle East. A qualitative literature review was conducted, analyzing peer-reviewed articles, policy reports, and gray literature from multiple regions. The study used a comparative approach to assess the cultural, political, and economic contexts shaping AI perceptions. It found significant regional differences in AI concerns: Western countries emphasize privacy, surveillance, and ethics, while regions like Africa and Asia focus on technological dependency, state control, and socio-economic issues, such as job displacement. Brazil and Argentina were noted for their AI governance models, while South Africa and Kenya leverage AI to tackle socio-economic challenges. The findings underscore the need for inclusive AI governance frameworks that respect diverse local contexts and values. These frameworks would contribute to a better understanding of AI's global impact and inform future policy development.**

**Correspondence:** papers@team.qeios.com — Qeios will forward to the authors

## 1. Introduction

The disparity in understanding the global perception of AI technologies is a critical issue, primarily stemming from the predominant emphasis on industrialized nations and Western viewpoints in current literature. The landscape of AI ethics exhibits a Western-centric bias, underscoring the necessity of integrating diverse perspectives to avoid reinforcing existing global power dynamics[1]. Recent research, including studies by Zhang and Dafoe[2], reveals that AI governance frameworks often neglect cultural and regional variations, potentially marginalizing the voices of non-Western societies. For instance, AI is usually perceived as a technological advancement and an economic development catalyst in Asia. In contrast, regions such as Africa and Latin America are influenced by historical contexts, including colonial legacies, which shape public apprehensions about the risks of AI perpetuating exploitation[3][4]. These findings highlight the urgent need for more inclusive AI frameworks that consider local socio-political, economic, and cultural factors. This aligns with Jobin et al.[1] advocacy for global AI ethics that authentically reflect diverse, non-Western values. Overall, the discourse surrounding artificial intelligence reveals significant concerns across various cultural and regional contexts. However, the prevailing literature largely mirrors Western perspectives, often overlooking the rich diversity of viewpoints from non-Western societies and cultures[2].

To create a fair approach to AI research and governance, it is crucial to include non-Western viewpoints and consider the socio-economic, political, and cultural factors that influence AI perceptions in different regions. Scholars like Sindermann et al.[5] and Kuziemski and Misuraca[6] emphasize the importance of local values and conditions in AI governance, especially in public sector decisions. Addressing these gaps is essential for developing inclusive AI governance frameworks that acknowledge regional differences and challenges.

## 2. Why It Matters

Rapidly implementing AI technologies in public services to improve efficiency could worsen current power imbalances and may not consider the specific socio-economic situations of various communities[6]. Government approaches to AI vary significantly by region. In more authoritarian regimes, AI apprehension is often tied to concerns about how the state will use AI to increase control over its citizens, mainly through mass surveillance technologies[7]. In more democratic societies, the apprehension is more centered around the ethical uses of AI in the private sector, such as preventing corporations from abusing personal data[8]. For instance, in China, the state's role in developing and deploying AI for surveillance and social credit systems heightens concerns about individual privacy. However, it does not provoke the same level of public debate seen in Western countries due to differences in political and cultural attitudes towards state intervention. In Europe, the General Data Protection Regulation (GDPR) has significantly addressed public apprehension about AI and privacy, leading to a more regulatory-focused discourse on AI risks[9].

Another critical area is understanding the varying levels of technological adoption, regulatory frameworks, and ethical concerns across different societies, which is crucial for addressing global AI challenges[10]. The ethical concerns surrounding AI differ significantly across regions, shaped by local cultural norms, governance structures, and socioeconomic conditions. These differences influence how societies perceive AI and regulate and adopt it. For example, countries with robust regulatory frameworks, like the European Union, prioritize data privacy and human rights concerns. At the same time, in Asian regions, the focus is more on regulations and economic impact[11] and more on innovation, seeing AI as a tool for accelerating technological progress[12].

Furthermore, the extent to which AI is embraced can significantly impact societal trust in these technologies. Tjilen et al.[13] emphasized that in regions with limited digital literacy and access to technology, there tends to be more fear and doubt surrounding AI. In contrast, in more technologically advanced societies, the focus of public discussion may shift towards regulatory measures rather than fear of the unknown. Therefore, considering AI from a cultural and regional perspective demonstrates the necessity of developing tailored approaches for AI governance and policy-making, considering diverse viewpoints. AI governance strategies should take into consideration cultural variations and local contexts. What may be effective in one region may be less successful in another, so policymaking should prioritize flexibility and adaptability[14].

## 3. Current And Common Apprehensions

Artificial intelligence (AI) concerns arise primarily from uncertainties regarding its long-term effects. Bostrom[15] asserts that the rise of superintelligent AI could present existential threats by surpassing human control and leading to unpredictable outcomes. This anxiety is closely linked to the "control problem," highlighting our inability to guarantee that AI systems align with human values. Furthermore, there is significant worry about potential job displacement. Scholars like Autor, Levy, and Murnane[16] illustrate how previous technological advancements, especially automation and AI, have triggered economic upheavals and reduced job opportunities. This has sparked fears of job losses[17], ethical dilemmas[18], and challenges related to data governance, particularly concerning corporate and government control[17] and bias issues[19].

Regrettably, while anxieties about AI differ across regions, most literature predominantly reflects Western viewpoints. In Western countries, worries typically center on job displacement, privacy breaches, and existential risks posed by superintelligent AI[15]. This prioritization echoes a broader discourse rooted in democratic ideals and individual rights, with ethical issues like algorithmic bias and accountability taking center stage[8]. European nations have responded with regulations like the General Data Protection Regulation (GDPR), showcasing their commitment to stringent ethical frameworks for privacy and data protection[9].

Conversely, non-Western regions often emphasize varying dimensions of AI's effects, shaped by unique political structures and cultural values. In Asia, for instance, AI is predominantly viewed as a driver of economic growth and technological supremacy. Nations like China perceive AI through a state-driven lens, leveraging it to augment national strength, particularly in surveillance and social credit systems[7]. This perspective frequently downplays privacy issues in favor of national security and efficient governance, starkly contrasting the Western focus on individual rights. While some surveillance-related concerns exist in China, they tend to be less emphasized due to cultural attitudes favoring state intervention[12].

In the Middle East, AI is often regarded as a means to bolster state authority and economic diversification. Countries like Saudi Arabia and the UAE are investing heavily in AI to advance national economic objectives and enhance surveillance capabilities, raising concerns about potential authoritarian governance[7]. Unlike Western discourse that leans toward issues of corporate data misuse and surveillance, Middle Eastern concerns about AI ethics often revolve around maintaining political stability and social order[8].

In Africa, fears surrounding AI are closely linked to the continent's colonial past and ongoing economic disparities. Research indicates that AI technologies might worsen socio-economic inequalities, given that Western companies predominantly design and control these systems. This leads to worries about "data colonialism" and the unequal distribution of AI's advantages[3][4]. Many African nations fear AI will exacerbate their technological dependency on the West, stifling local innovation opportunities and deepening existing divides[20].

The risk of AI replacing low-skilled jobs is particularly alarming, especially given the continent's high unemployment and poverty rates[3]. However, there are potential benefits; AI is being leveraged to tackle crucial challenges in sectors like poverty alleviation and healthcare, especially in nations like South Africa and Kenya, where it is viewed as a tool for social good[20][21].

These regional disparities underscore the complexities and nuances of global AI apprehension. As AI development accelerates, inclusive frameworks incorporating diverse cultural perspectives and effectively addressing the specific concerns of non-Western societies are urgently needed.

# 4. Methodology

### *Choice of Countries (Non-Western)*

These countries were selected for their leadership in AI adoption and policy development. They are engaged in significant local and global challenges related to the ethical use of AI, privacy concerns, and socio-economic impacts. For instance, South Africa is prioritizing data governance[20], Kenya is engaged in discussions about the socio-economic ramifications of AI[21], and Brazil is tackling issues related to privacy, citizen rights, and government accountability in the implementation of AI.

- Middle East (Saudi Arabia, UAE, Egypt)
- Africa (Kenya, South Africa)
- Asia (Japan, China)
- Latin America (Argentina, Brazil)

### *4.1. Comparative Thematic Analysis*

A thematic analysis was performed to identify key themes in existing literature, focusing on privacy concerns, job displacement, state surveillance, ethical governance, and economic impact across regions. Themes were coded to compare Western and non-Western contexts.

Data Collection:

The primary data for this analysis were derived from a comprehensive literature review of academic articles, policy documents, and reports from organizations focusing on AI governance, perceptions, and cultural differences across regions (e.g., Africa, Asia, Latin America, and the Middle East).

Relevant studies were gathered from databases such as Google Scholar, JSTOR, and ScienceDirect using keywords such as "AI apprehension," "AI governance," "AI ethics," and "regional AI perceptions."

Peer-reviewed articles and gray literature were included to ensure a wide range of perspectives.

Coding Process:

- The coding process began by reviewing all selected sources to identify common themes related to AI apprehension. These themes were grouped into **privacy concerns, job displacement, government surveillance**, **economic impact**, and **ethical issues.**
- The analysis revealed specific sub-themes, such as **AI for social good**, **data sovereignty**, **and AI in governance,** based on regional focus.

A systematic coding process was used to categorize the extracted data. Key terms, phrases, and concepts from each article were coded to identify regional patterns of AI apprehension. These codes were organized into broader themes like "privacy concerns," "state surveillance," and "economic impact of AI." The iterative coding process allowed for revisions of themes and codes based on continuous literature analysis, ensuring that insights from non-Western perspectives were included in the comparative framework.

Thematic Categories:

The literature findings were organized into broader categories that matched the conceptual framework. These categories included:

- **Political Governance:** Explore how different political regimes (democratic vs. authoritarian) affect AI governance and public perceptions regarding state control and data protection.
- **Economic Context:** Assess the socio-economic factors, including concerns about job displacement and increased economic inequality due to AI.
- **Cultural Norms:** Investigate how cultural values influence public attitudes toward AI, focusing on trust in technology and its effects on social structures.

Cross-Regional Comparison:

The thematic analysis highlighted regional differences in AI apprehension. In Western regions, concerns were mainly about **data privacy** and ethical implications for **individual rights.** In contrast, authoritarian states like China focused more on **state surveillance** and **technological control.** This approach enabled the research to compare regional differences and commonalities in AI perceptions while identifying specific local concerns requiring customized governance frameworks.

Using Comparative Thematic Analysis, the study sought to establish a systematic approach to understanding how political, economic, and cultural factors shape AI perceptions in different regions, providing a deeper understanding of global AI apprehension.

### *4.2. PRISMA Flow Diagram*

The PRISMA (Preferred Reporting Items for Systematic Reviews and Meta-Analyses) flow diagram is a standard tool for showing how studies are selected and included in systematic reviews and meta-analyses. It visually represents the steps in identifying, screening, and evaluating studies for inclusion.

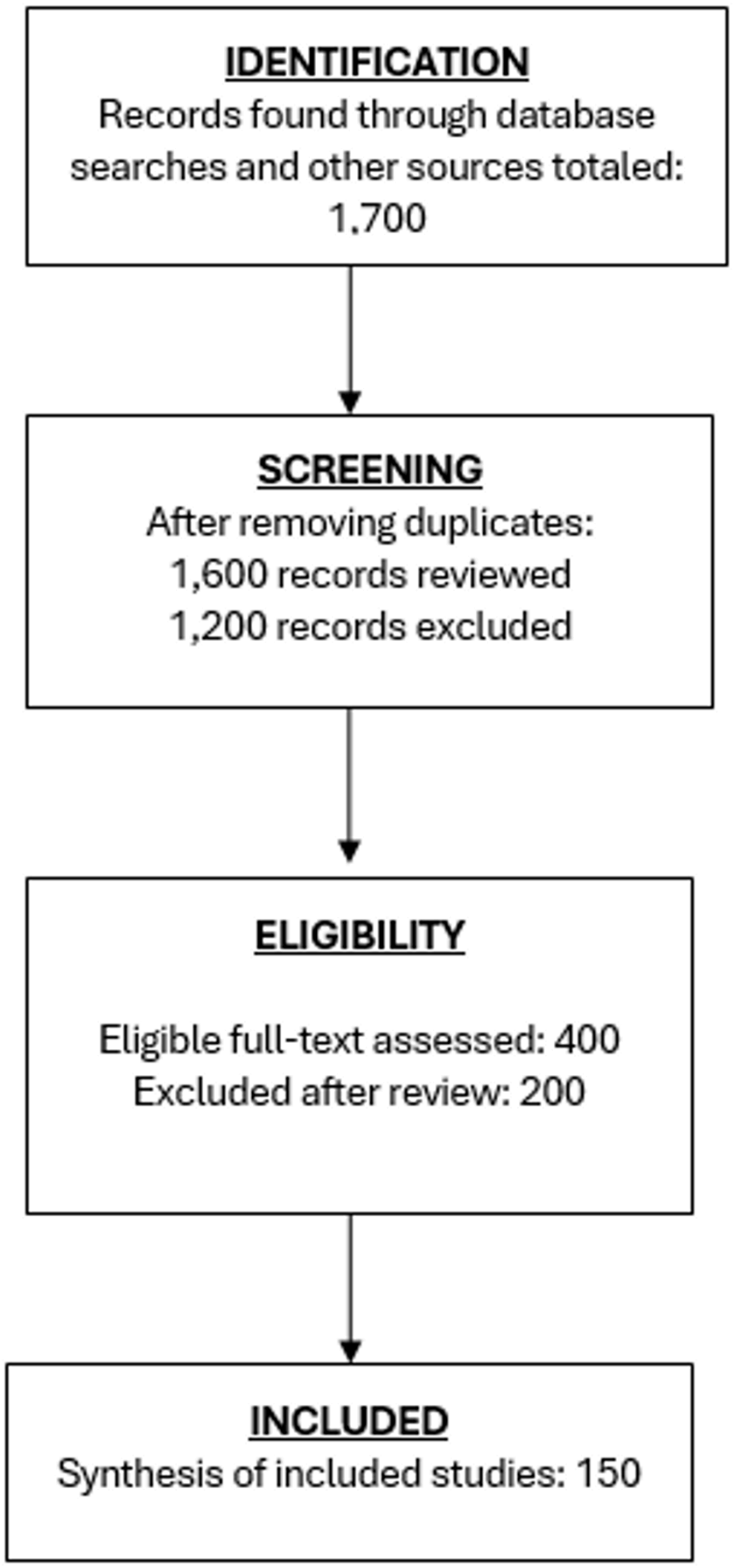

**Figure 1.** *PRISMA Flow*

The PRISMA flow diagram shows the study's selection process, indicating the number of records at each stage.

1. Identification:
    - A total of 1,500 studies were initially found through database searches on JSTOR, Google Scholar, and ScienceDirect using keywords like "AI apprehension," "cultural perspectives on AI," "regional AI governance," and "AI ethics in non-Western societies."
    - The researcher's proficiency is restricted to English, affecting the identification and inclusion process.

- Additional records were found from other sources, including think tanks, government agencies, and international organizations: 200 records.
- Total Records Found: 1,700

2. Screening:
   - Total records after removing duplicates: 1,600
   - Records screened: All 1,600 based on title and abstract
   - Records excluded: 1,200, as they did not focus on regional AI perceptions, governance, or ethical concerns
   - Records for Full-Text Review: 400
3. Eligibility:
   - Full-text articles evaluated for eligibility: 400 articles were assessed based on inclusion criteria related to regional and cultural apprehensions surrounding AI.
   - Studies excluded: After a thorough review of the complete text, 200 studies were excluded due to insufficient empirical data or a focus on unrelated themes, such as technical AI models lacking a cultural or regional perspective.
   - Studies to Include in the Final Review: 200
4. Inclusion:
   - A total of 150 studies were included in the qualitative synthesis based on geographic focus, publication date (2018-2023), and relevance to the thematic analysis criteria.

### *4.3. Consistency*

A second coder, knowledgeable in AI governance and regional studies, was hired to code a subset of the literature. Comparing the initial and second rounds of coding ensured inter-coder reliability and maintained a consistent application of themes and categories across sources.

Data Triangulation: The analysis used various sources, including academic articles, reports from international organizations, and gray literature like government reports and think tank publications. This method ensured that findings were validated and minimized biases from any source.

Using a structured and systematic approach to coding, thematic analysis, and comparative evaluation, the study provides a comprehensive understanding of the diverse regional perspectives on AI apprehension, ensuring depth and consistency in the analysis.

## 5. Conceptual Framework

Figure 2 outlines the structure for understanding how political, economic, and cultural factors influence AI apprehension in different regions. Each main category includes sub-factors that elaborate on the reasons driving regional concerns about AI.

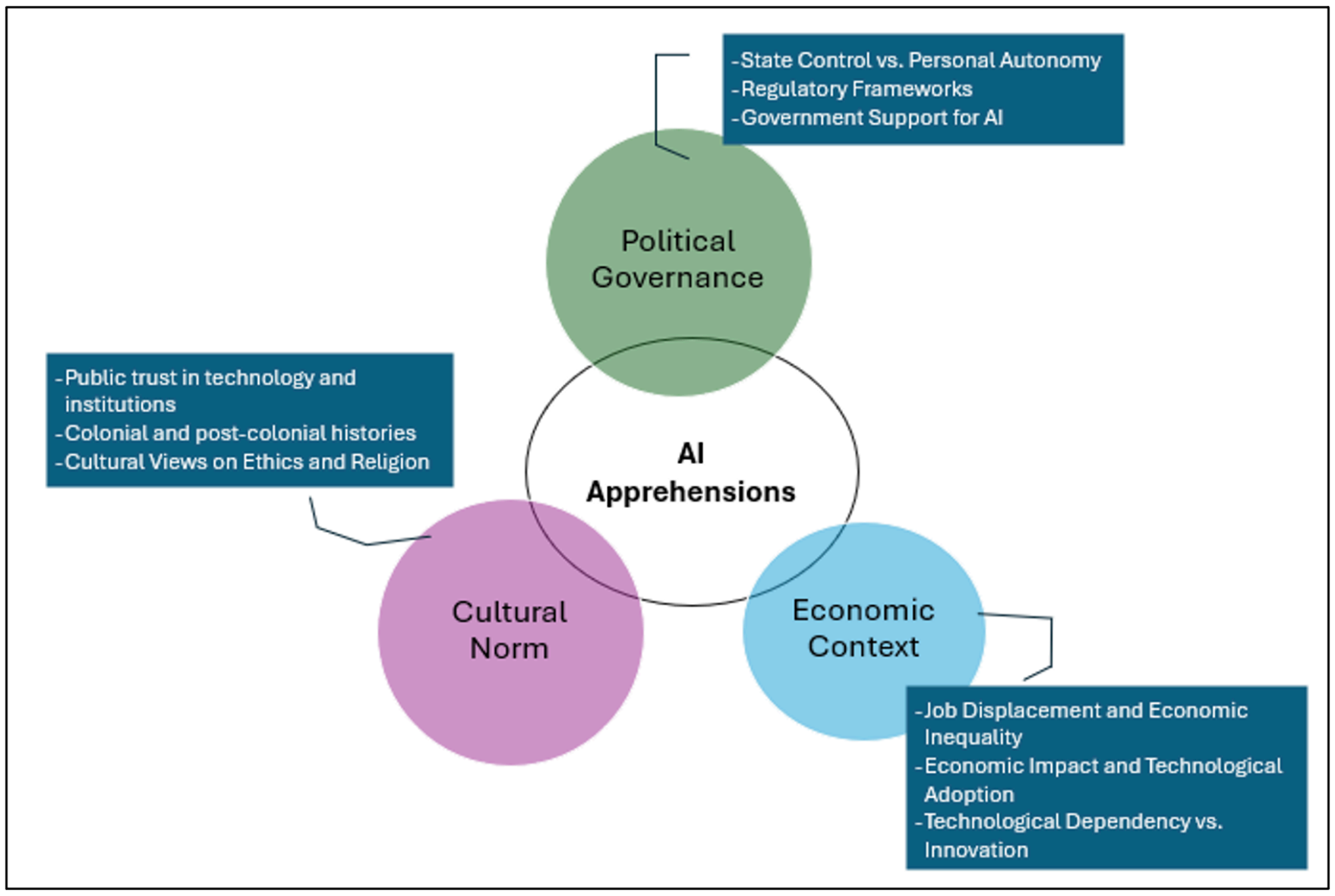

**Figure 2.** *Conceptual Framework for AI Apprehension*

The central theme focuses on the concerns and uncertainties surrounding AI's use and impact across different regions, influenced by socio-cultural, political, and economic factors.

**Political Governance** examines how political systems and regulatory measures impact the societal effects of AI, highlighting the need for a balance between state control and individual freedom within regulatory frameworks.

**Economic Context** delves into how issues such as job displacement, inequality, and the adoption of AI shape public perceptions of technology and its socio-economic implications.

**Cultural Norms** investigate how cultural attitudes toward technology, trust in institutions, historical contexts, and ethical values influence the acceptance and integration of AI within society.

Figure 2 allows for a comparative analysis of AI governance and its societal impacts in the selected regions, which will be discussed next. This analysis focuses on common concerns and different perspectives and strategies for addressing AI governance in each area.

## 6. Findings

Table 1 presents a comprehensive overview of the critical concerns, regional differences, and cultural influences that affect the perception and governance of artificial intelligence (AI) across various regions. The analysis identifies both patterns and contradictions in these perspectives. For instance, Western nations often prioritize the ethical implications and privacy issues associated with AI, whereas in authoritarian regimes, the focus shifts toward the potential of AI to enhance state control through surveillance technologies[7][12]. Furthermore, in Africa, historical colonial legacies and apprehensions regarding technological dependency significantly shape the attitudes toward AI. Conversely, in South

Africa and Kenya, AI is increasingly recognized as a valuable tool for advancing social good[3][21].

| Region/Country | Key AI Apprehension/Challenges | Cultural/Social Factors |
|---|---|---|
| **Western Nations** | Concerns about AI include its impact on employment, privacy, existential risks, and ethical issues like algorithmic bias and accountability. Key focus areas are regulation, data protection (e.g., GDPR), and individual rights. | Significant concerns exist about individual rights, privacy, and job displacement. Ethical standards are emphasized in the development of AI. |
| **Middle East (Saudi Arabia, UAE, Egypt)** | Concerns are rising about the authoritarian use of AI for state control and surveillance. AI monitoring citizens' online activities threatens freedom of expression. | Political structure affects public perception—AI is viewed through the lens of state control, surveillance, and security. |
| **Africa (General)** | Concerns include the deepening of socioeconomic divides, job displacement due to automation, and the digital divide. AI is feared to exacerbate existing inequalities and reinforce technological dependency on Western nations.<br>**South Africa:** AI improves society by reducing poverty through tools like AI-driven maps and advancing healthcare. However, there are concerns about foreign companies controlling local data.<br>**Kenya:** AI helps improve areas like agriculture, finance, and urban planning. However, there are concerns about how it is governed, especially regarding AI surveillance. Issues include privacy violations and the risk of too much government control. | Colonial legacies shape concerns of exploitation and unequal access to AI benefits.<br>**South Africa:** Focuses on local data control to keep AI benefits within the local economy.<br>**Kenya:** Balancing innovation with data protection and privacy is essential, especially given weak enforcement capacities. |
| **Asia (Japan, China)** | **Japan:** Positive perception of AI is mainly due to the cultural integration of robots and AI into everyday life. AI is seen as an extension of human labor, not a threat. Limited apprehension about job displacement as AI is viewed as complementary to human work, especially in elder care.<br>**China:** AI is seen as a tool for technological leadership and state control. Due to the political system, there is less emphasis on privacy concerns, though some reservations remain about surveillance and state control. | **Japan:** Cultural environment emphasizes harmony, societal benefit, and human-robot collaboration.<br>**China:** Focus on technological progress and state control, with less concern about individual privacy. |
| **Latin America (General)** | AI apprehension focuses on privacy, government surveillance, and mass data collection. Concerns about AI being used to suppress dissent and violate human rights, particularly in politically unstable countries like Venezuela. | Trust in government and perceptions of political corruption influence AI acceptance. |

| Region/Country | Key AI Apprehension/Challenges | Cultural/Social Factors |
|---|---|---|
| | **Argentina:** Focus on ethical AI utilization, transparency, and non-discrimination in AI governance. Promoting public involvement in policy formation.<br>**Brazil:** Privacy concerns and government surveillance, particularly in fragile democracies. Ethical concerns regarding AI systems exacerbate inequalities and marginalize communities: gender gap and indigenous language concerns in AI tools. | **Argentina:** Commitment to human rights in AI development and inclusive governance strategies.<br>**Brazil:** Distrust of AI use in government surveillance, concerns over fairness and accountability, and the cultural impact of AI systems on marginalized communities. |

**Table 1.** *Key Findings of AI Apprehension*

### *6.1. Regional Variations in AI Apprehension*

In Western nations, there is significant concern about AI's impact on employment, privacy, and potential existential risks. Ethical challenges, including algorithmic bias and AI accountability, are also prominent in these regions, emphasizing regulation, data protection (e.g., GDPR in Europe), and the potential for AI to infringe on individual rights[8].

In the Middle East, concerns about AI are shaped by the region's authoritarian political structures and the rapid pace of AI adoption for state control. In Saudi Arabia and the UAE, AI is being aggressively integrated into national strategies for economic diversification, particularly in reducing reliance on oil. However, there is growing apprehension about using AI for surveillance, mainly as these countries invest heavily in AI-driven cybersecurity systems. This has led to concerns about individual privacy and the role of AI in enhancing state power. In Egypt, for example, public discourse has emerged around the use of AI in monitoring citizens' online activities, with critics warning that these technologies could suppress freedom of expression and increase state surveillance.

In contrast, Japan has a relatively positive perception of AI, shaped by the cultural integration of robots and AI into everyday life. Japan's history with robotics has created a cultural environment where AI is seen as an extension of human labor rather than a threat. For instance, AI is being used in elder care facilities to support the aging population, and there is relatively little apprehension about job displacement, as AI is perceived to complement human workers rather than replace them. This cultural attitude towards technology fosters a more optimistic view of AI and its potential to enhance societal harmony, a key value in Japanese society.

In China, AI is regarded as a tool for technological leadership and state control, with comparatively less emphasis on privacy concerns due to differences in the political system. Nonetheless, there are still reservations about AI's use for surveillance and state control.

In Africa, AI apprehension is deeply intertwined with the continent's history of colonialism and ongoing economic inequalities. For instance, Alonso et al. 2000[22] highlight how AI technologies may deepen existing socio-economic divides, with AI development concentrated in the hands of a few elite technologists, leaving the majority without access to its benefits. In some African countries, AI is often viewed through the lens of its impact on employment, as automation threatens to replace low-skill jobs that are vital for large segments of the population[3]. Artificial intelligence is making poverty reduction possible by improving the collection of poverty-related data through poverty maps[20]. The digital divide in these regions exacerbates public concern, as many fear AI will worsen existing inequalities[3]. Additionally, there are concerns that AI could reinforce patterns of technological dependency on Western nations, a legacy of colonialism[3].

### *6.1.1. South Africa: AI for Social Good and Data Sovereignty*

South Africa has positioned itself as a leading proponent of AI adoption on the continent. The government has underscored the pivotal role of AI in propelling social and economic development, mainly through initiatives that harness AI to address poverty, improve healthcare, and expand educational outreach. For instance, AI is instrumental in creating comprehensive poverty maps that inform policy decisions related to resource allocation, directly contributing to poverty alleviation efforts[20]. Additionally, AI is leveraged to analyze extensive healthcare datasets, enabling the prediction of disease outbreaks and enhancing the allocation of medical resources. These applications vividly illustrate AI's potential to confront critical social challenges within the region.

The issue of data sovereignty has become increasingly prominent in national discussions. With the growing presence of foreign technology companies in Africa's AI sector, there are concerns that African nations may cede control over their data, potentially leading to a type of "data colonialism"[4]. In response, South Africa's AI governance frameworks have implemented policies mandating international companies to collaborate with local businesses and adhere to stringent data protection regulations. These policies aim to ensure that the benefits of AI advancement remain within the local economy and that local data is safeguarded against external exploitation.

### *6.1.2. Kenya: Balancing Innovation with Ethical Governance*

Kenya is one of the countries where AI is being incorporated into national development strategies, particularly in sectors such as agriculture, financial services, and urban planning. The Kenyan government has promoted AI as a catalyst for innovation, as seen in initiatives like "AgriTech," which harnesses AI to enhance farming practices, boost crop yields, and alleviate food insecurity. Furthermore, AI is increasingly utilized in mobile banking platforms to provide financial services to underserved populations, thus promoting inclusion[21].

However, Kenya needs to work on fostering innovation and upholding ethical governance. The deployment of AI-powered surveillance technologies in urban areas of Nairobi has raised worries about privacy breaches and the possibility of excessive government control. Although Kenya's Data Protection Act of 2019 was a significant step toward addressing these concerns by setting out guidelines for data privacy, challenges persist in enforcing it due to limited institutional capacity and technical know-how. These challenges underscore the delicate balance between promoting AI innovation and ensuring that AI governance frameworks safeguard the rights and privacy of citizens.

### *6.1.3. Brazil: AI Scrutiny*

In Latin America, AI apprehension often focuses on privacy concerns and government surveillance. In countries with fragile democracies or high levels of political corruption, such as Venezuela, the use of AI for mass surveillance has heightened fears that these technologies will be used to suppress dissent and violate human rights. Public trust in AI, therefore, is closely linked to perceptions of government accountability and transparency.

A case study from Brazil shows how AI-driven surveillance technologies have been used in urban areas to fight crime. However, there are concerns about civil liberties and the potential for government abuse of these systems. In recent years, Brazil has made progress in AI governance and aims to prioritize responsible and ethical AI governance as a core part of its vision for the future. The economic impact of the AI market in Brazil is projected to increase from around $3 billion in 2023 to $11.6 billion by 2030, with a GDP impact of 6-8%. The country's strengths lie in its data policies and e-participation.

However, there are concerns that these AI systems could exacerbate existing inequalities by disproportionately affecting marginalized communities, sparking ethical concerns about fairness and accountability. Additionally, there is a significant gender gap in STEM education. With 274 indigenous languages, there is a risk of these languages being marginalized in data sets and AI tools, potentially leaving many indigenous language speakers behind, especially in the case of large language models.

### *6.1.4. Argentina: National AI Plan*

Argentina has adopted a unique approach to AI governance, prioritizing ethical AI utilization and citizen data protection. Launched in 2020, Argentina's National AI Plan is designed to foster responsible AI development while upholding human rights. The country has established principles for the ethical application of AI, focusing on transparency, accountability, and non-discrimination across both public and private sectors. Argentina's AI governance framework also promotes public involvement in AI policy formation, positioning the country as a regional trailblazer in developing inclusive and transparent AI strategies.

# 7. Cultural Factors

## *7.1. Cultural Factors Shaping AI Apprehension*

Studies indicate that in societies with higher levels of institutional trust, where people trust their governments and institutions to regulate AI effectively, there is less fear about the consequences of AI[2].

### *7.1.1. Trust in Technology and Institutions*

The level of public trust in technology and institutions significantly influences the approach to AI governance in different countries, as depicted in Table 1. Studies show that trust in government plays a crucial role in shaping public perceptions of e-government services and, by extension, AI governance. For example, Horsburgh et al.[23] emphasize that the trustworthiness of governmental institutions is vital for gaining public support for e-government initiatives, similar to the importance of trust in AI systems for their acceptance and effective governance[23]. This connection underscores that without a basis of trust, efforts to implement AI technologies may encounter significant public resistance.

Furthermore, the research by Zhang and Kim[24] indicates that public trust in government can be shaped by perceptions of government performance, especially in the context of corruption. Their study suggests that citizens' trust is influenced by immediate government actions and long-term perceptions of governance quality, which can impact how AI governance is viewed and embraced[24]. This highlights the significance of ethical governance and transparency in fostering public trust in AI systems, as citizens are more inclined to support AI initiatives when they believe their government operates with integrity and accountability. In addition, Yousaf et al.'s findings underscore that the government's unethical conduct can erode public trust, which is crucial for the effective implementation of AI governance frameworks[25]. This correlation is further reinforced by the insights of Winfield and Jirotka, who contend that ethical governance is vital for cultivating trust in AI and robotics. They suggest that a lack of ethical considerations can lead to public skepticism and resistance[26]. Therefore, the interaction between public trust in institutions and the governance of AI technologies is intricate and multifaceted, requiring a meticulous approach that prioritizes transparency, accountability, and ethical standards.

#### *7.1.2. Religious and Ethical Perspectives*

Religious and ethical perspectives play a crucial role in shaping AI governance, mainly as these factors influence societal norms and expectations regarding technology. Integrating ethical considerations into AI governance frameworks is essential for fostering public trust and ensuring the responsible development and implementation of AI systems. For instance, Winfield and Jirotka emphasize that ethical governance is fundamental to building trust in robotics and AI systems, proposing a roadmap that links ethics, standards, regulation, and public engagement as critical components of effective governance[26]. This framework highlights the necessity of incorporating diverse ethical viewpoints, including religious perspectives, to address the multifaceted challenges posed by AI technologies.

Moreover, the governance of AI must also consider the implications of religious diversity and the interactions between religious and non-religious actors in public policy. Martínez-Ariño discusses how local governance networks can facilitate the regulation of public concerns, including those related to technology, by incorporating the voices of various stakeholders, including religious organizations[27]. This approach underscores the importance of recognizing and integrating religious and ethical perspectives into the governance of AI, as these perspectives can significantly influence public acceptance and the ethical deployment of AI systems.

The global landscape of AI ethics guidelines, as articulated by Jobin et al.[1], reflects the diverse interests of various stakeholders, including religious groups, in shaping the ethical frameworks that govern artificial intelligence. The engagement of different organizations in establishing AI principles signifies a collective recognition of the necessity for ethical guidance that resonates across various cultural and religious contexts. Thus, understanding and integrating religious and ethical perspectives into AI governance is crucial for ensuring that AI technologies are developed and implemented in ways that honor cultural values and promote social justice. The Universal Guidelines on AI and the UNESCO Recommendations on the Ethics of AI provide comprehensive international frameworks to address the ethical implications of artificial intelligence. The Universal Guidelines advocate for unified ethical principles in AI development, emphasizing transparency, fairness, the protection of privacy, and accountability. Similarly, the UNESCO Recommendations on the Ethics of AI, adopted in 2021, offer a more detailed framework that underscores the importance of inclusivity, human dignity, transparency, justice, fairness, non-maleficence, responsibility, and the safeguarding of fundamental rights in AI governance[1].

### *7.1.3. Colonial and Post-Colonial Legacies*

In regions with a history of colonial exploitation, particularly in Africa and Latin America, there is concern that AI technologies may replicate colonial-era patterns of resource extraction and dependency (Table 1). Birhane[4] critically analyzes how contemporary AI technologies may perpetuate colonial dynamics, especially in African contexts. Birhane argues that the motivations behind algorithmic practices mirror those of historical colonialism, emphasizing the corporate-driven nature of modern exploitation. This raises concerns about replicating dependency and resource extraction patterns reminiscent of the colonial era. This perspective is crucial for understanding the socio-political implications of AI in regions with a legacy of colonialism. For instance, in South Africa, AI governance frameworks have increasingly focused on protecting local data and ensuring that AI development benefits local economies. The government has implemented policies that require international companies to partner with local businesses and adhere to local data protection laws when deploying AI technologies[20]. Munn[28] addresses the intersection of digital labor and AI, emphasizing how these technologies can perpetuate exploitative practices reminiscent of colonial resource extraction and underscore the urgent need to confront the colonial logic embedded within contemporary AI systems, particularly in regions historically subjected to exploitation. Nikalje and Çiftçi[29] provide insight into the psychological ramifications of colonial mentality, which can be extrapolated to understand the apprehensions surrounding AI technologies in post-colonial contexts. By illustrating how colonial attitudes perpetuate feelings of inferiority and dependency among marginalized groups, their research underscores the potential for AI to replicate these historical patterns of exploitation and reinforce existing inequalities in regions like Africa and Latin America. This connection highlights the importance of critically examining the socio-cultural implications of AI deployment in historically colonized societies.

In the realm of AI governance within post-colonial regions, it is essential to create AI systems that are cognizant of local biases and do not perpetuate historical injustices. Areas with legacies of systemic exclusion or marginalization —such as caste systems in South Asia, racial inequities in Latin America, or ethnic disparities in Africa—necessitate thorough bias audits of AI models. Without these evaluations, AI technologies risk exacerbating existing inequalities. AI systems must be developed with a deep understanding of local socio-cultural contexts and subjected to regular testing to ensure they do not reinforce the discriminatory practices that have historically oppressed vulnerable communities[4][3]. A pertinent example includes using AI in recruitment processes, facial recognition, and criminal justice systems. These systems can reinforce biases if not meticulously managed, as they often mirror the biases present in their training data[19]. In addition to technical audits, involving cultural leaders, religious scholars, and community elders in AI policy discussions is advantageous. Their involvement can help ensure that AI systems are aligned with local ethical principles and societal values. By incorporating diverse perspectives from these cultural and religious figures, AI governance frameworks can be enriched with nuanced ethical considerations that resonate with the community. This collaboration not only aids in identifying culturally significant values but also fosters community acceptance and builds trust in AI technologies. These leaders can offer valuable insights regarding the ethical boundaries and social expectations that should guide AI development and implementation, ensuring that AI governance remains responsive to local contexts and community needs[1][27]. Engaging these stakeholders also addresses concerns about "data colonialism" and external entities' potential exploitation of local communities[4].

### *7.1.4. Societal Attitudes Toward Employment and Automation*

A study by Gursoy and Chi[30] delves into the impact of cultural attitudes on the acceptance of artificial intelligence (AI) devices in the tourism industry. The research underscores the importance of considering cultural context when examining the incorporation of AI technologies in various sectors, including hospitality and airline services. The findings indicate that tourists' readiness to embrace AI varies across services. Moreover, the study suggests that cultural attitudes toward work and employment can offer valuable insights for shaping broader AI governance strategies.

For instance, in China, the government has positioned AI as a crucial driver of economic expansion, placing less emphasis on concerns about job displacement[31]. This reflects a firm conviction in the potential of AI to enhance productivity and technological innovation, with government-led initiatives propelling AI development across different sectors. Conversely, apprehensions about job displacement have been a focal point of public discourse on AI in Western Europe and North America. Policies in these regions underscore the necessity of social safety nets, retraining programs, and labor market protections to alleviate potential adverse effects of AI on employment[32]. This divergence in governance approach is rooted in cultural attitudes toward work and social welfare.

Elamin and Omair[33] also provide insights into how cultural attitudes, particularly those related to gender roles, can shape perceptions of work and employment within specific socio-cultural contexts. The study reveals that traditional attitudes toward working women persist among Saudi males, with variations influenced by age and education. This underscores the broader implications of cultural attitudes on labor dynamics and governance, particularly in the context of automation and AI.

## 8. Study Limitations and Future Research Directions

Artificial intelligence is evolving quickly, with frequent developments and shifts in public opinion. As a result, current literature may soon need to be updated.

This highlights the need for ongoing research to track how regional perceptions of AI change over time, especially as the technology becomes more widespread and its impact on daily life increases.

The study primarily uses literature from peer-reviewed journals, books, and reports published in the last five years. While this approach highlights recent trends, it may lead to selection bias, as some regions or viewpoints might be underrepresented. Countries with less academic research or visibility could need more adequate representation, potentially skewing the understanding of AI apprehension. Additionally, focusing on English-language sources may limit diverse perspectives, particularly from regions where local research is published in other languages.

The findings of this study provide valuable insights into regional perceptions of AI but may only apply to some countries in the discussed regions. The selected countries were chosen based on existing literature, yet each region has significant variation in socio-economic conditions, political governance, and cultural norms. For instance, attitudes toward AI in advanced economies like Japan and South Korea may differ from those in lower-income countries such as India and Indonesia. Therefore, the study's conclusions on AI apprehension might not be universally applicable, and further research is needed to understand these differences more thoroughly.

This study is based on a qualitative literature review. It does not include primary empirical data, such as surveys or interviews, which could offer deeper insights into individuals' experiences and perceptions in different regions. As a result, the conclusions drawn from secondary data may only partially capture the changing dynamics of AI perceptions across various areas.

## 9. Conclusion

This article underscores the pressing need to expand the worldwide conversation on artificial intelligence (AI) by integrating diverse cultural, regional, and socio-political viewpoints. The existing literature on AI concerns has predominantly focused on Western perspectives, overlooking the distinct challenges and issues faced by non-Western societies. Through a comparative examination of AI concerns in regions such as Asia, Africa, Latin America, and the Middle East, this study emphasizes the significance of localized factors — including historical legacies, political systems, and socio-economic conditions — in shaping public perceptions and governance of AI technologies. The findings underscore the necessity for adaptable and context-sensitive AI governance. Culturally attuned policies, ethical standards, and regulatory frameworks are imperative for addressing the specific needs and worries of diverse populations. A uniform approach to AI governance risks perpetuating global disparities and amplifying existing socio-economic gaps. Pursuing empirical research that complements this qualitative analysis by collecting primary data through surveys, interviews, or case studies in non-Western regions is crucial. Further exploration is needed to understand how particular socio-cultural factors influence the deployment of AI technologies in diverse settings. Collaboration among policymakers, researchers, and local stakeholders is essential for crafting AI governance models that prioritize inclusivity, fairness, and cultural resonance. Expanding the global AI discourse will ensure more equitable outcomes and enhance the societal advantages of AI technologies worldwide.

## Statements and Declarations

### *Data Availability*

The data for this review article consist of publicly available literature sourced from academic databases (JSTOR, Google Scholar, ScienceDirect) and gray literature as described in the methodology. Specific search terms and the full list of included and excluded studies based on the PRISMA process are available from the author upon reasonable request.

### *Author Contributions*

Conceptualization: V.F.; Methodology: V.F.; Investigation: V.F.; Writing—Original Draft Preparation: V.F.; Writing—Review & Editing: V.F. (Victor Frimpong). The author has read and agreed to the published version of the manuscript.

## Declarations

**Funding:** No specific funding was received for this work.

**Potential competing interests:** No potential competing interests to declare.